\documentclass[twoside,11pt]{article}

\usepackage{jmlr2e}
\usepackage{amsmath}

\ShortHeadings{}{Deep Learning in Medicine}
\firstpageno{1}

\begin{document}

\title{Experimenting with Knowledge Distillation techniques for
performing Brain Tumor Segmentation}

\author{\name Ashwin Prakash Nalwade \email  \\
       \addr Courant Institute of Mathematical Sciences\\
       New York University\\
       New York, NY 10012, USA
       \AND
       \name Jacqueline Kisa \email  \\
       \addr Vilcek Institute of Graduate Biomedical Sciences \\
       New York University\\
       New York, NY 10012, USA}

\editor{}

\maketitle

\begin{abstract}%   <- trailing '%' for backward compatibility of .sty file
Multi-modal magnetic resonance imaging (MRI) is a crucial method for analyzing the human brain. It is usually used for diagnosing diseases and for making valuable decisions regarding the treatments - for instance, checking for gliomas in the human brain. With varying degrees of severity and detection, properly diagnosing gliomas is one of the most daunting and significant analysis tasks in modern-day medicine. Our primary focus is on working with different approaches to perform the segmentation of brain tumors in multi-modal MRI scans.  Now, the quantity, variability of the data used for training has always been considered to be crucial for developing excellent models. Hence, we also want to experiment with Knowledge Distillation techniques.
\end{abstract}


\section{Background}
A glioma is a type of tumor known as \textit{intra-axial}, as they initiate from glial cells within the brain and often mix with normal brain tissues. As one of the most common cancers, it is increasingly important to diagnose it accurately and quickly to decrease mortality rates following diagnosis. 
\newline
\newline
Multi-modal magnetic resonance imaging (MRI) is a crucial method for analyzing the human brain. It is usually used for diagnosing diseases and for making valuable decisions regarding the treatments - for instance, checking for gliomas in the human brain. Therefore, the accuracy in the assessment of MRI results is paramount, which requires expertise, time, and focus. Lack of any of these could lead to unsatisfying outcomes. Typically, these scans are analyzed by clinical experts, so it puts restrictions on the amount of data available for making decisions - therefore it was inevitable that a lot of research effort at the intersection of medical and computer science has been attempted in addressing this issue, and thus it was of interest to us too.

\section{Motivation and Related Work}
 Gliomas are the pillar of numerous different studies focusing on their heterogeneous nature. With varying degrees of severity and detection, properly diagnosing gliomas is one of the most daunting and significant analysis tasks in modern-day medicine. To do this more accurately, researchers have incessantly concentrated their efforts to research better methods for performing Brain Tumor Segmentation. 
\newline
\newline
Our primary focus is on working with different excellent approaches to perform the segmentation of brain tumors in multi-modal MRI scans [\cite{isen}, \cite{cascade},  \cite{myro}].  Now, the quantity, variability of the data used for training has always been considered to be crucial for developing excellent models. Especially in the context of medical imaging - since for many areas there isn't a easy way to acquire huge volumes of data, as opposed to other computer vision based tasks, such as segmenting objects [food, landmarks, automobiles] or people. Hence, we also want to explore the relationship between model performance and the amount of data utilized for training. We plan to do this by leveraging knowledge distillation \citep{hinton2015distilling} techniques \citep{10.1007/978-3-030-46643-5_32}. 
\newline
\textbf{HYPOTHESIS}: Models leveraging knowledge distillation techniques will outperform the original stand-alone models.

\section{Data}
The BraTS dataset \citep{bench} is curated every year by MICCAI and is publicly available for download at the Center for Biomedical Image Computing and Analytics (CBICA) website at the University of Pennsylvania, on the SICAS Medical Image Repository [SMIR], and is also hosted on Kaggle. The training dataset consists of 259 High-Grade Gliomas (HGG) and 76 Low-Grade Gliomas (LGG), 66 Gliomas for validation, and 191 Gliomas for testing. Each MRI scan describes four modalities - native (T1), post-contrast T1-weighted (T1Gd), T2-weighted (T2), and T2 Fluid Attenuated Inversion Recovery (T2-FLAIR) volumes. All BraTS multi-modal scans are available as NIfTI files (.nii.gz format), and we make use of the nibabel library for processing these scans. Different modalities of magnetic resonance imaging have the capability to indicate tumor-induced tissue changes from different perspectives - bringing about variability in the types of representations we can learn, and hence this is beneficial for the brain tumor segmentation task when these modalities are processed collectively.  

\section{Evaluation metrics}
The dice coefficient (DICE), also called the dice score or overlap index is a very frequently used metric utilized for segmentation tasks \citep{dice}, and specifically, the validation of segmentation tasks in medical imaging. In the most basic terms, we can define the dice score as twice the area of intersection between target sample and predicted sample divided by the sum of the areas of the target sample and the predicted sample.  

\section{Knowledge Distillation}
We make use of the model distillation \citep{radso} technique (which is a sub-category of knowledge distillation) and follow the workflow adopted by \cite{10.1007/978-3-030-46643-5_32}. An example is provided in the figure below. There are various techniques which can be used for ensembling [like majority ensemble], but we chose averaging of the model outputs for the ensemble.

\begin{figure}[htp]
    \centering
    \includegraphics[width=6cm]{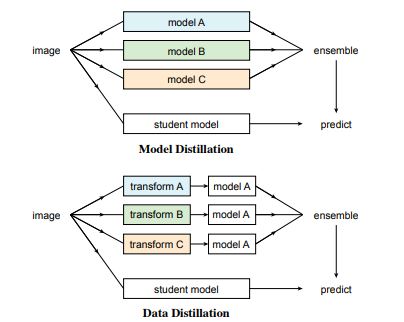}
    \caption{Distillation techniques. For model distillation, we would use the ensemble of multiple models for annotating unlabeled data, and train the student model on the original data plus the previously unlabeled data. The objective is for the student model to have higher predictive power than the original models}
    \label{}
\end{figure}

\section{Methodology}

During the pre-processing of data, for each input image, all input volumes [modalities] are normalized to achieve unit variance and zero mean for non zero foreground voxels. During the data augmentation step, we adopt widely practiced techniques like image cropping, image rotation, scaling, mirroring. Moreover, for each modality, contrast and intensity shift augmentations are applied.  
\newline
For the deep learning part, we used 3D UNet and its variants [Residual UNets and Cascaded UNets]. These architectures and their specific structures/layers are chosen on the basis of approaches which have been used by the previous winners of various iterations [years] of the BraTS challenges. The first step is to train multiple models by themselves on the dataset and compare performances - from this, we can recognize the best performing stand-alone model. Further, we would create an ensemble of the models to annotate the unlabeled data and train a distilled model on that newly annotated data in addition to the original data [model distillation process]. Note that by unlabeled data, we mean the half of the validation data and twenty percent of the test data that we have set aside and consider it as unlabeled data. The labeling process of MRI's requires highly trained medical experts, so it is very difficult or even impossible to retrieve trustable unlabeled data for this task - for this reason, we have to reconfigure and make changes in the train, validation, and test sets themselves. Then, we would use the best performing stand-alone model to train on this new combination of original training data and annotated data, and compare performances with the previous models.

\subsection{UNet}
Backed by empirical evidence, strong theoretical foundations, and a large number of research works, UNets \citep{unet} are undeniably considered as one of the most efficient architectures used to address the BioMedical Image Segmentation problem. It consists of an encoder and a decoder for learning representations, and encoder networks have the capability to efficiently capture highly abstract and higher level features w.r.t MRI images [by doubling the number of feature maps after each level]. We make use of the UNet architecture adopted by \citet{isen} since it allowed them to achieve 2nd place in the BraTS 2018 challenge. Isenee et al. make a number of modifications in order to get excellent results, such as replacing rectified linear unit [ReLU] activations with Leaky ReLU and leveraging trilinear upsampling with the decoder. They also decided to use instance normalization instead of batch normalization as it has been well documented that the latter does not perform [learn] well with small batch sizes. Before training this network, we use a patch size of 128x128x128 down from the original which is 240x155x155. We set the batch size as two and set the initial learning rate as 2x$10^{-4}$ and a decay rate of 0.60. For the loss, a combination of Soft Dice Loss and Binary Cross-Entropy was used [unweighted sum of both the losses], and we perform the segmentation task of segmenting three regions - Enhancing Tumor [ET], Tumor Core [TC] and Whole Tumor [WT]. The reason why we do not rely on just dice loss alone is that 
while being widely popular and providing state of the art results on many medical segmentation challenges, the dice loss has some downsides, such as more instances of error-ridden softmax probabilities and convergence problems when compared to negative log likelihood [cross-entropy loss].   
The loss functions are given as:

\begin{align*}
L^{dice}(g,p) = \frac{1}{K} \cdot \sum_{k=1}^{K} \frac{2 \sum p_k \cdot g_k}{\sum p_k^{2} + g_k^{2}}
\end{align*}
\begin{align*}
L^{bce}(g,p) = -\frac{1}{N} \cdot \sum_{k=1}{K} \sum (g_k \cdot log(p_k) + (1-g_k) \cdot log(1 - p_k))\\
 \end{align*}
 
 where N represents the number of output voxels, K is the number of regions [3 in this case], $g_k$ is the ground truth and $p_k$ is the prediction for a region k. Thus, the overall loss is given by:
 \begin{align*}
     L^{total}(g,p) = L^{dice}(g,p) + L^{bce}(g,p)
 \end{align*}
 
 \subsection{Cascaded Unet}

We adopt the Cascaded UNet used by \citet{cascade} which achieved the 14th position in the BraTS challenge. Cascaded UNet consists of multiple blocks that learn representations associated with each modality in a separate manner. ReLU activation is used as a non-linearity, and the network contains basic residual blocks having two ReLU layers, two instance normalization layers and two convolutional layers. For training, we adopt SGD optimizer along with a starting learning rate of 0.1, decay rate of 0.85, and batch size equal to 4.

\begin{figure}[!tbp]
  \centering
  \begin{minipage}[b]{0.9\textwidth}
    \includegraphics[width=\textwidth]{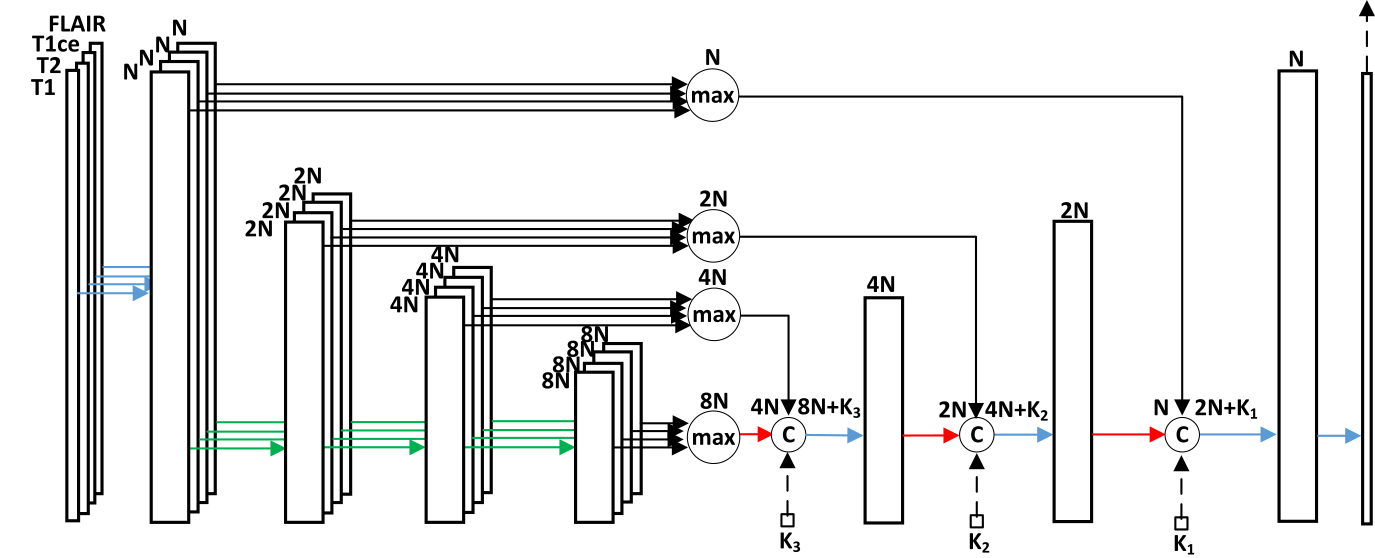}
    \caption{Cascaded UNet Architecture}
  \end{minipage}
  \hfill
  \begin{minipage}[b]{0.1\textwidth}
    \includegraphics[width=3cm]{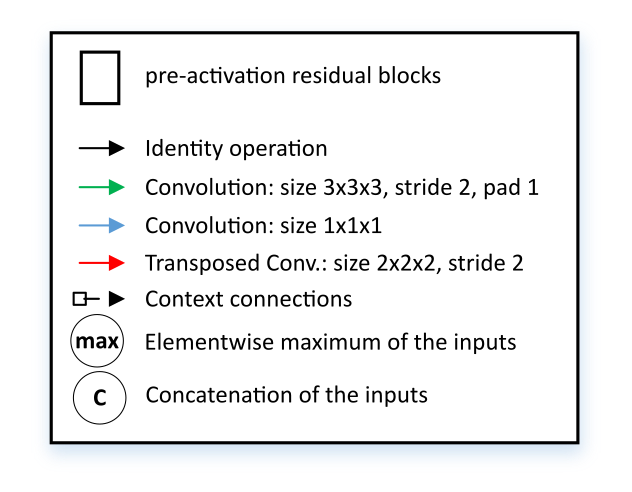}
    %\caption{Index}
  \end{minipage}
\end{figure}

\subsection{Residual Unet}
Residual networks \citep{resid} are widely recognized for learning identity mappings well and also for mitigating the vanishing gradient problem. Due to the skip connections, they are able to achieve a smooth flow of gradients. They also have a smaller training time associated with them compared to networks with comparable power and allow the features learned from the previous layers to be passed forward in an efficient manner. \citet{myro} are of the same opinion, and they added residual blocks to UNet to improve performance. We adopt their architecture for training the residual UNet. \citet{myro} also utilize group normalizations, as leveraging it as a normalization layer has been previously showcased to give excellent results even with smaller batch sizes used during training. Furthermore, the calculations related to group normalization are also independent of the batch size, so we can get similar performance over varying sizes. The ReLU function is used for non-linearity, and they also use a variational auto-encoder part for regularizing effect. The model is trained with the same learning rate and decay as the UNet model, using the Adam optimizer. \citet{myro} use a patch size of 144x144x128, however we choose to keep the same patch size as was used for the UNet. The training for all the models was performed for 280 epochs.

\section{Results}
We report the dice scores obtained for different models across different tumor types.
\begin{table}[h]
\centering
\begin{tabular}{|l|l|l|l|}
\hline
Approach & Enhancing Tumor & Tumor Core  & Whole Tumor  \\ \hline
UNet    & 0.71871  & \underline{0.81343}   & 0.84591 \\ \hline
\underline{Residual UNet} & \underline{0.72585}  & 0.80872  & \underline{0.86415}  \\ \hline
Cascaded UNet & 0.72197  & 0.81129   & 0.85654 \\ \hline
Ensemble & 0.74916 & 0.81733 & \textbf{0.87682}  \\ \hline
Distilled Model & \textbf{0.75187}  & \textbf{0.82661} & 0.87074  \\\hline
\end{tabular}
\caption{\label{tab:table-name}Dice scores on the test set. Underlined values represent optimal scores for the stand-alone models, and the values in bold represent the globally optimal scores}
\end{table}

We can see from the table that Residual UNet achieves better dice scores than UNet and Cascaded UNet for Enhancing Tumor and Whole Tumor, hence Residual UNet was the better performing stand-alone model. Thus, it was chosen for the distillation process. Considering the global optimal scores, we observe that the distilled model achieved better scores than the ensemble and stand-alone models  for Enhancing Tumor and Tumor Core.

\section{Analysis}
The distilled model performed better than the ensemble model for the Enhancing Tumor and Tumor Core categories, and attained very close performance to the ensemble for the Whole Tumor. 
\newline
Compared to \citet{10.1007/978-3-030-46643-5_32}, our distilled performed better in two categories as compared
to better performance in one category in theirs - however, their dice scores were better
overall. One of the reasons for this difference is that we trained for a lesser number of epochs than them, so longer training could increase the performance.  
\newline
As for alternative strategies, instead of using the best performing stand-alone model for knowledge distillation, we could have once again used an ensemble to train on newly annotated data in addition to the
original data. We could have also performed more extensive hyper-parameter tuning and run cross validation tests, and as a result we might have achieved more optimal dice scores. All this was not feasible due to time constraints. Keeping all these factors in mind, our hypothesis was successfully validated for two tumor types. The extension of our work would definitely involve the validation of the hypothesis for all three types along with some significant improvement in the dice scores at the same time.
\newline
Recently, researchers have proposed a novel framework, named TumorGAN \citep{tgan}, to generate image
segmentation samples by leveraging the unpaired adversarial training method. The results in the paper
[verified on the BraTS dataset] showcase that the synthetic data samples generated by their proposed method can considerably improve performance for tumor segmentation when applied to
segmentation network training. This could be a direction worth pursuing in the future.
\newline
Contribution - Jacqueline worked on the UNet and Residual UNet and Ashwin worked on the Cascaded UNet, Ensemble, and Distilled model.

% Acknowledgements should go at the end, before appendices and references

% Manual newpage inserted to improve layout of sample file - not
% needed in general before appendices/bibliography.

% uncomment this
%\newpage

%\appendix
%\section*{Appendix A.}
%\label{app:theorem}

% Note: in this sample, the section number is hard-coded in. Following
% proper LaTeX conventions, it should properly be coded as a reference:

%In this appendix we prove the following theorem from
%Section~\ref{sec:textree-generalization}:

%\bibliography{sample}

\end{document}